\documentclass{article}

\PassOptionsToPackage{numbers, compress}{natbib}

\usepackage[final]{neurips_2023_ml4ps}

\usepackage[utf8]{inputenc} 
\usepackage[T1]{fontenc}    
\usepackage{hyperref}       
\usepackage{url}            
\usepackage{booktabs}       
\usepackage{amsfonts}       
\usepackage{nicefrac}       
\usepackage{microtype}      
\usepackage{xcolor}         
\usepackage{amsmath}
\usepackage{amssymb}
\usepackage{mathtools}
\usepackage{amsthm}
\usepackage{subcaption}
\usepackage{graphicx}

\title{Towards out-of-distribution generalization in large-scale astronomical surveys: robust networks learn similar representations}

\author{%
  Yash Gondhalekar$^1$\\
  \texttt{yashgondhalekar567@gmail.com} \\
  \And
  Sultan Hassan$^{2,3,4,5}$\\
  \texttt{sultan.hassan@nyu.edu} \\
  \And
  Naomi Saphra$^{6,7}$\\
  \texttt{nsaphra@fas.harvard.edu} \\
  \And
  Sambatra Andrianomena$^{8,9}$\\
  \texttt{andrianomena@gmail.com} \\
\AND
$^1$\normalfont{BITS Pilani, K.K. Birla Goa Campus} \quad 
$^2$New York University \quad
$^3$Flatiron Institute \\
$^4$University of the Western Cape \quad
$^5$NASA Hubble Fellow \\
$^6$Kempner Institute for the Study of Natural and Artificial Intelligence  \quad
$^7$Harvard University \\
$^8$South African Radio Astronomy Observatory \quad
$^9$University of the Western Cape
}

\begin{document}

\maketitle

\begin{abstract}
    The generalization of machine learning (ML) models to out-of-distribution (OOD) examples remains a key challenge in extracting information from upcoming astronomical surveys. Interpretability approaches are a natural way to gain insights into the OOD generalization problem. We use Centered Kernel Alignment (CKA), a similarity measure metric of neural network representations, to examine the relationship between representation similarity and performance of pre-trained Convolutional Neural Networks (CNNs) on the CAMELS Multifield Dataset. We find that when models are robust to a distribution shift, they produce substantially different representations across their layers on OOD data. However, when they fail to generalize, these representations change less from layer to layer on OOD data. We discuss the potential application of similarity representation in guiding model design, training strategy, and mitigating the OOD problem by incorporating CKA as an inductive bias during training. 
\end{abstract}

\section{Introduction}
Although the astronomy and cosmology communities have embraced ML methods, many key challenges remain. We focus on two such goals: the interpretability of ML models and their robustness under distribution shifts. 

Robustness is particularly salient in astronomy because simulations only provide an approximate realization of the observed universe. In addition, different simulation models provide different realizations of the same astrophysical observables, so models trained on simulations from one environment may not generalize to simulations from another. A model that fails under a distribution shift between simulations may also fail when provided with real-world data, so it is crucial to assess whether a model is robust to a given distribution shift. A better understanding of the settings under which a model fails to generalize may guide us toward better OOD generalization. Better interpretations of ML models may also allow us to discover new hidden features within trained models, which hold significant importance in astronomy \cite{2021arXiv211114566N}.

Recently, astronomy research has begun to apply interpretability techniques from machine learning. \citet{PhysRevD.102.123506} used saliency methods to interpret deep learning models trained to recover cosmological parameters from weak lensing maps. \citet{10.1093/mnras/sty2575} used latent tree structures to analyze the relationship between model performance and data. \citet{Wu_2020} used the Gradient-weighted Class Activation Mapping attribution tool to interpret the connection between galaxies' morphological features and gas content. \citet{2023arXiv230501582C} proposed using symbolic regression to discover mathematical expressions that approximate neural networks.


\citet{2022arXiv220609868C} applied CKA to study robustness to adversarial examples, and we similarly consider the relationship between robustness and similarity in our setting. Specifically, we compare similarities of the internal representations of pre-trained CNNs using CKA, finding that when models fail to generalize under distribution shift, they tend to produce representations that remain similar between layers. We then discuss the possible connection between representation similarities and accuracy and how these insights can be used to promote robustness under distribution shift, suggesting ways to improve model training or prune neural network architectures. 


\section{Methods}
\subsection{Data}
We use the publicly available CAMELS Multifield Dataset (CMD) \cite{CMD2021}\footnote{\url{https://camels-multifield-dataset.readthedocs.io/en/latest/index.html}}, which is an open-access collection of 2D maps and 3D grids of 13 different fields created using different hydrodynamic (IllustrisTNG--henceforth, {\sc TNG}, and {\sc SIMBA}) and pure $N$-body simulations as part of the CAMELS project \cite{CAMELS}. We here use 2D maps with $256\times256$ pixels and size $25\,  {\rm Mpc}/h$ of the total matter density (Mtot) and Gas temperature (Temperature) from the {\sc TNG} and {\sc SIMBA} simulations at $z = 0$. The total matter density constitutes baryonic and dark matter contributions. The CMD dataset contains 1,000 simulations with 15 distinct maps per simulation. 
\subsection{CKA similarity measure}\label{sec:cka}
CKA compares the representations produced by different layers of the same or different architectures on shared input data. CKA is the normalized version of the Hilbert-Schmidt Independence Criterion (HSIC), which is used to test the dependence on distributions. Such a normalization allows the CKA metric to be invariant under isotropic scaling of the representations. The steps to compute CKA (assuming a linear kernel) are as follows (see \citet{2019arXiv190500414K} for more details). The input to CKA is a pair of representations $X \in \mathbb{R}^{m \times n_X}$ and $Y \in \mathbb{R}^{m \times n_Y}$ where $m$ is the number of examples and $n_X, n_Y$ are the respective feature dimensions. These inputs are transformed into gram matrices denoted by $K$ and $L$ satisfying $K = XX^T$ and $L = YY^T$, such that $K$ and $L$ have shape $m \times m$. The resulting gram matrices are then centered using a centering matrix, $H = I_n - \dfrac{1}{n}11^T$, to produce $K' = HKH$, $L' = HLH$. Since CKA is the normalized version of the HSIC metric, the HSIC is first calculated by taking the dot product between the flattened versions of the centered gram matrices: $\textrm{HSIC}(K, L) = \dfrac{\textrm{vec}(K') \cdot \textrm{vec}(L')}{(m - 1)^2},$
where $\textrm{vec}$ transforms the matrix into a vector. The CKA is given by $\textrm{CKA}(K, L) = \dfrac{\textrm{HSIC}(K, L)}{\sqrt{\textrm{HSIC}(K, K) \textrm{HSIC}(L, L)}}.$
Following \citet{nguyen2021wide}, we use the mini-batch CKA approach to reduce computational expense.

\subsection{Implementation}\label{sec:implementation}

Our CKA implementation closely follows \citet{nguyen2021wide}. The CKA calculation is performed on publicly available pre-trained CNNs on the CMD datasets from \citet{2021arXiv210909747V}. The basic architecture of these pre-trained CNN models consists of a series of blocks of convolutional layers $\rightarrow$ BatchNorm $\rightarrow$ Leaky ReLU layers followed by two fully connected layers with dropout. These CNNs were trained to predict the six cosmological and astrophysical parameters.
We select the ``best'' trial based on validation loss, following the approach of \cite{Villaescusa-Navarro_2022} for calculating the CKA similarities using 50 maps for each field\footnote{The results remain unaffected using different numbers of examples as long as sufficient examples ($\gtrsim16-32$) are used.}. A forward pass of the pre-trained CNN is performed on the test set two times independently, and the CKA similarity of each layer's output representation (from the first pass) is computed for every layer's output representation (from the second pass), yielding a CKA matrix.

To test the possible correlation between CKA similarities and performance, we compute the coefficient of determination $R^2$ score between the estimated and the true parameters. We only focus on recovering the cosmological parameters, $\Omega_m$ and $\sigma_8$, since most CNNs can marginalize over the astrophysics.

\section{Results}
\label{sec:results} 
We will refer to train-test setting pairs by name, e.g., {\sc TNG--SIMBA} for the case where the CNN is trained on maps from {\sc TNG} and tested on maps from {\sc SIMBA}. We consider four cases to compute the CKA similarity matrices: {\sc TNG--TNG} and {\sc TNG--SIMBA} for Temperature maps, and {\sc SIMBA--SIMBA} and {\sc SIMBA--TNG} for Mtot.

\begin{figure*}[tp!]
  \begin{minipage}[t]{.48\linewidth}
  \begin{subfigure}[t]{\textwidth}
    \centering
    \includegraphics[width=\textwidth,keepaspectratio]{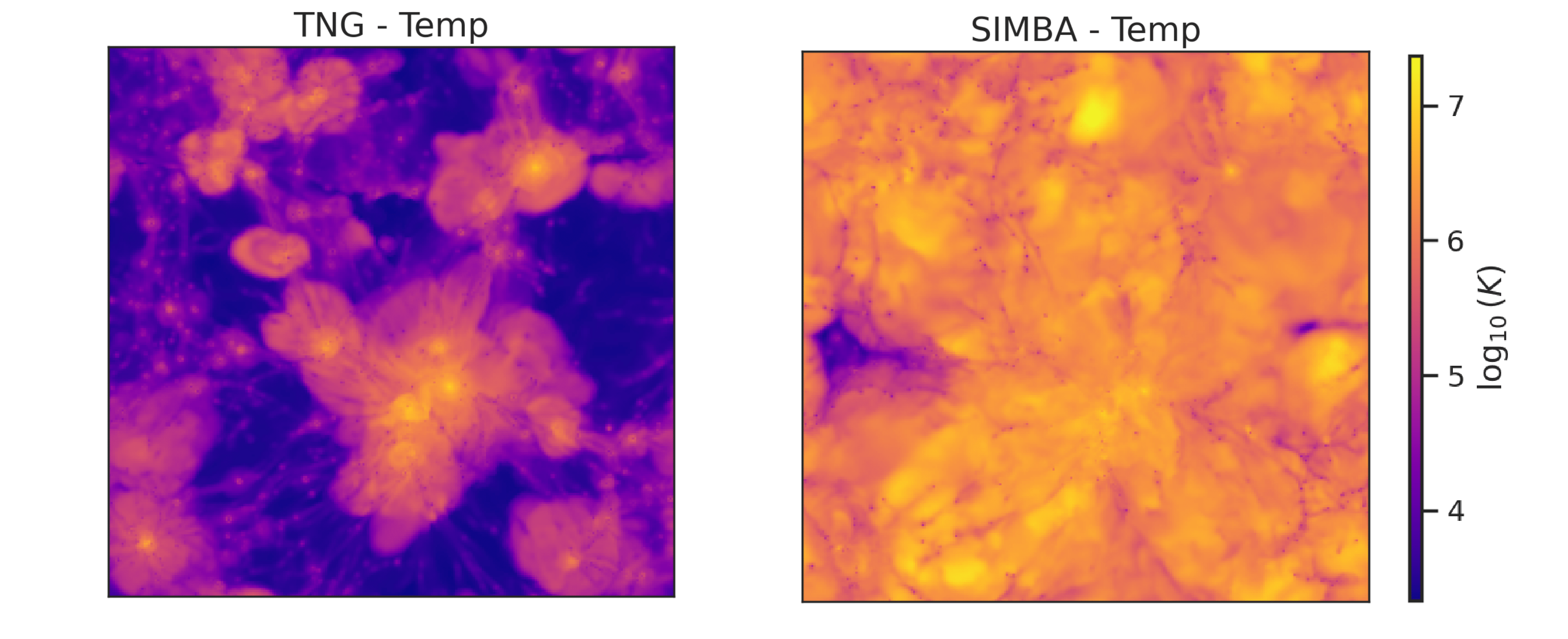}
    \caption{Random example of the Temperature field from the {\sc TNG} and {\sc SIMBA} simulations.}\label{fig:temp_ex}
  \end{subfigure}
  
  \vspace{3mm}
  \begin{subfigure}[t]{\textwidth}
    \centering
    \includegraphics[width=\textwidth,keepaspectratio]{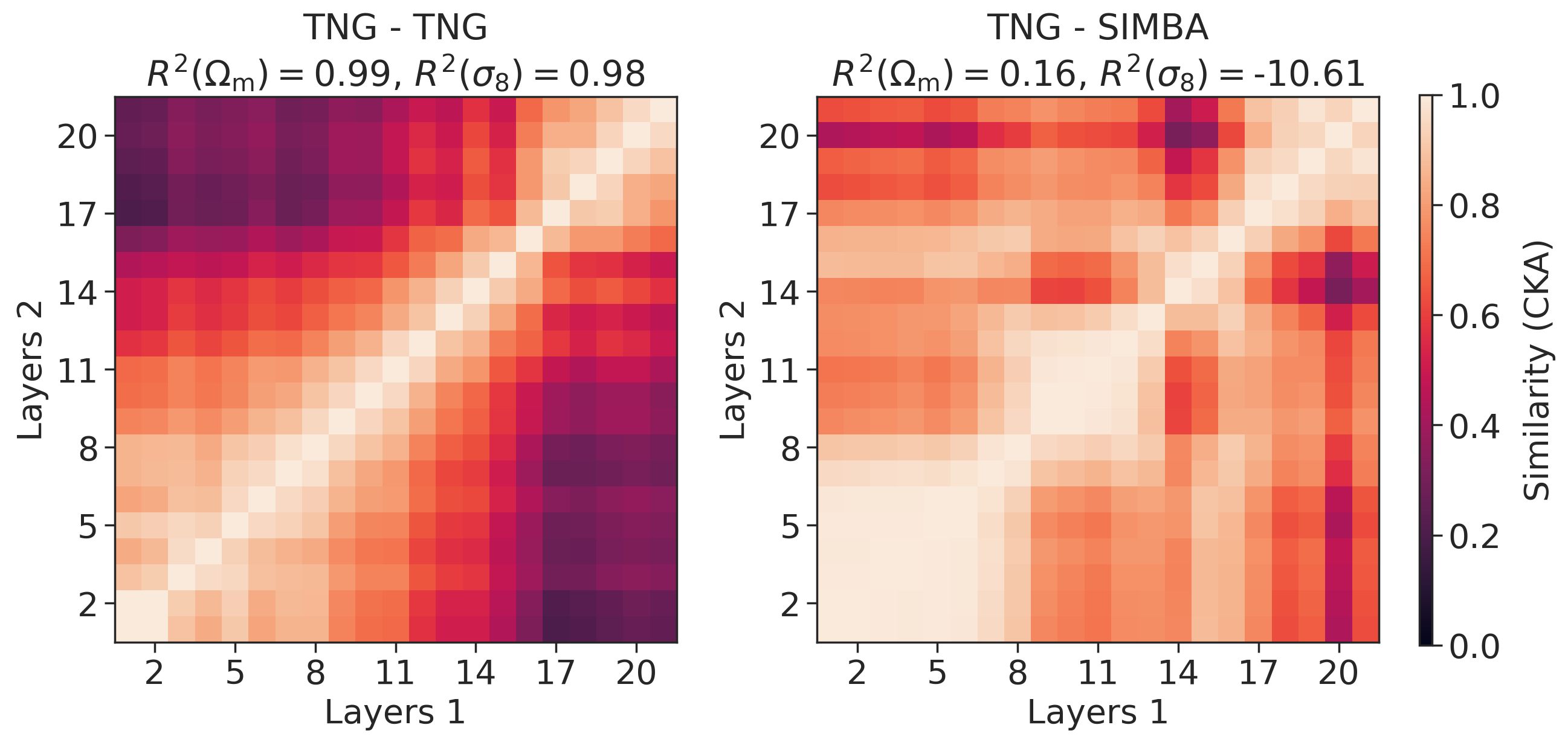}
    \caption{CKA matrices for the ID ({\sc TNG-TNG}) (left) and OOD ({\sc TNG-SIMBA}) (right) cases. The titles show $R^2$ scores between model prediction and true value for the $\Omega_{\rm m}$ and $\sigma_8$ cosmological parameters.}\label{fig:temp_cka}
  \end{subfigure}
  
  \vspace{3mm}
  \begin{subfigure}[t]{\textwidth}
    \centering
    \includegraphics[width=\textwidth,keepaspectratio]{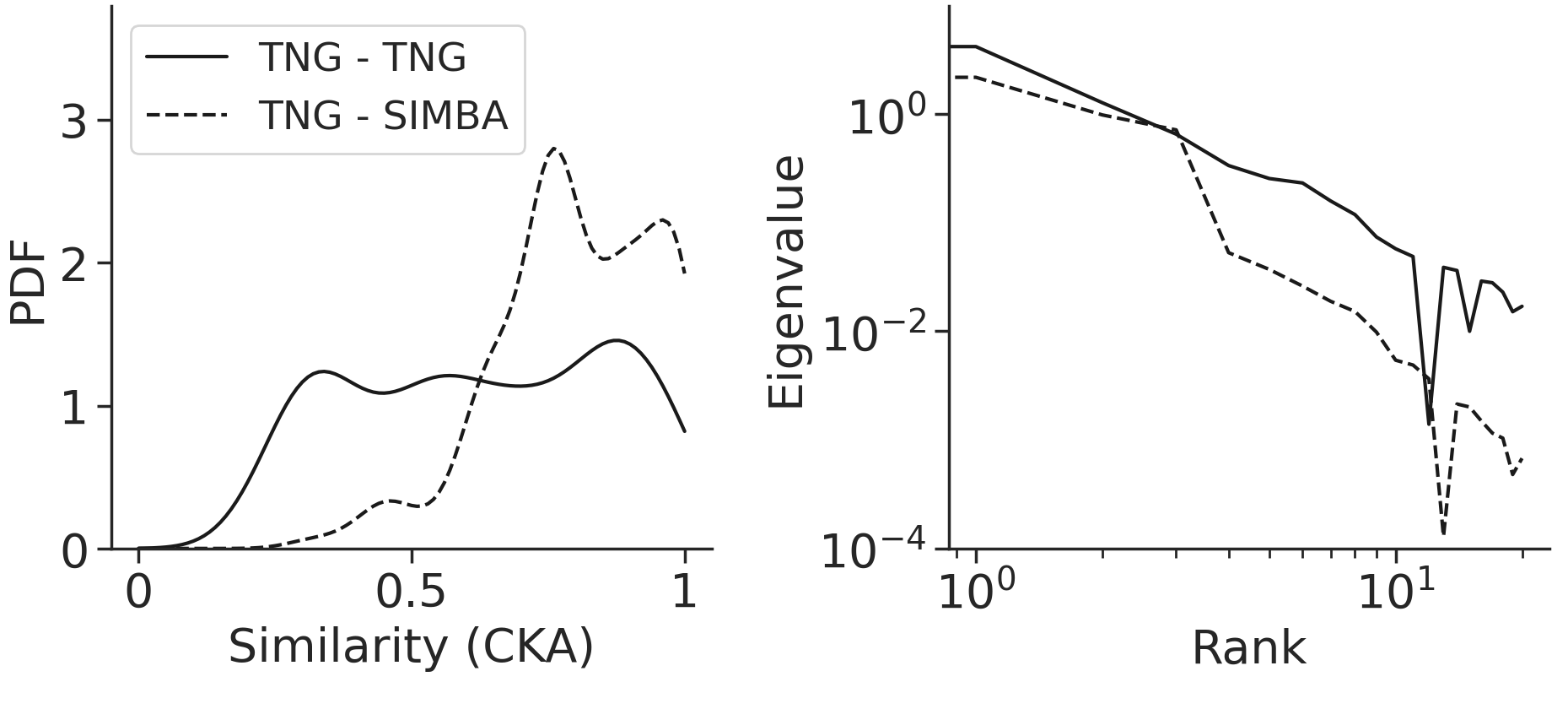}
    \caption{Summary statistics of the CKA matrices from Fig.~\ref{fig:temp_cka}. On the left is the probability density function (PDF) of the CKA similarities, and on the right is the eigenvalues as a function of rank (i.e., the spectrum).}\label{fig:temp_stats}
  \end{subfigure}
  \caption{Analysis of the CKA similarities for the Temperature field. Representations of layers of the CNN trained on the {\sc TNG} simulations are diverse only when the CNN is tested on 2D maps from {\sc TNG} (i.e., ID samples) but are almost stagnant on 2D maps from {\sc SIMBA} (i.e., OOD samples). The $R^2$ scores for {\sc TNG-SIMBA} are drastically inferior to those for {\sc TNG-TNG}, suggesting that the model fails to generalize to OOD ({\sc SIMBA}) data.}\label{fig:T}
  \end{minipage}\hfill
  \begin{minipage}[t]{.48\linewidth}
  \begin{subfigure}[t]{\textwidth}
    \centering
    \includegraphics[width=\textwidth,keepaspectratio]{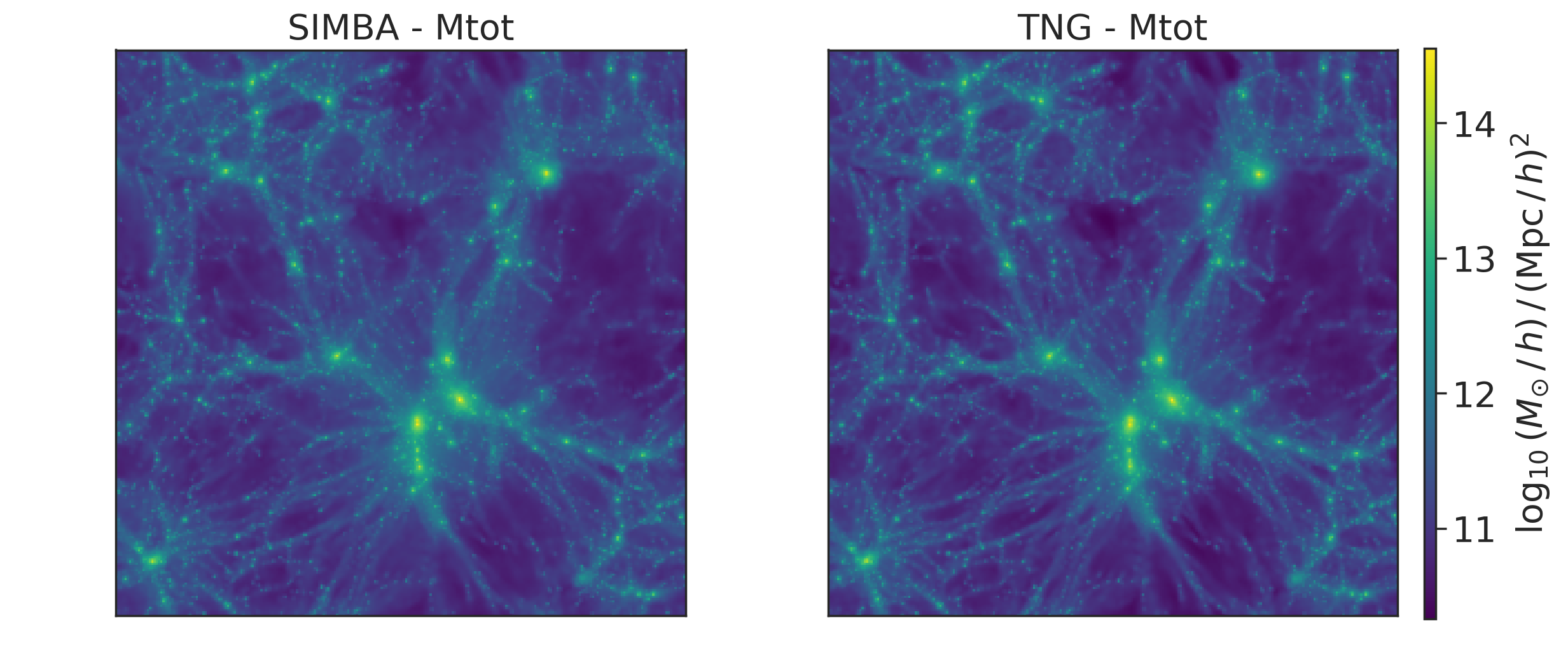}
    \caption{Random example of the Mtot field from the {\sc SIMBA} and {\sc TNG} simulations.}\label{fig:mtot_ex}
  \end{subfigure}
  
  \vspace{3mm}
  \begin{subfigure}[t]{\textwidth}
    \centering
    \includegraphics[width=\textwidth,keepaspectratio]{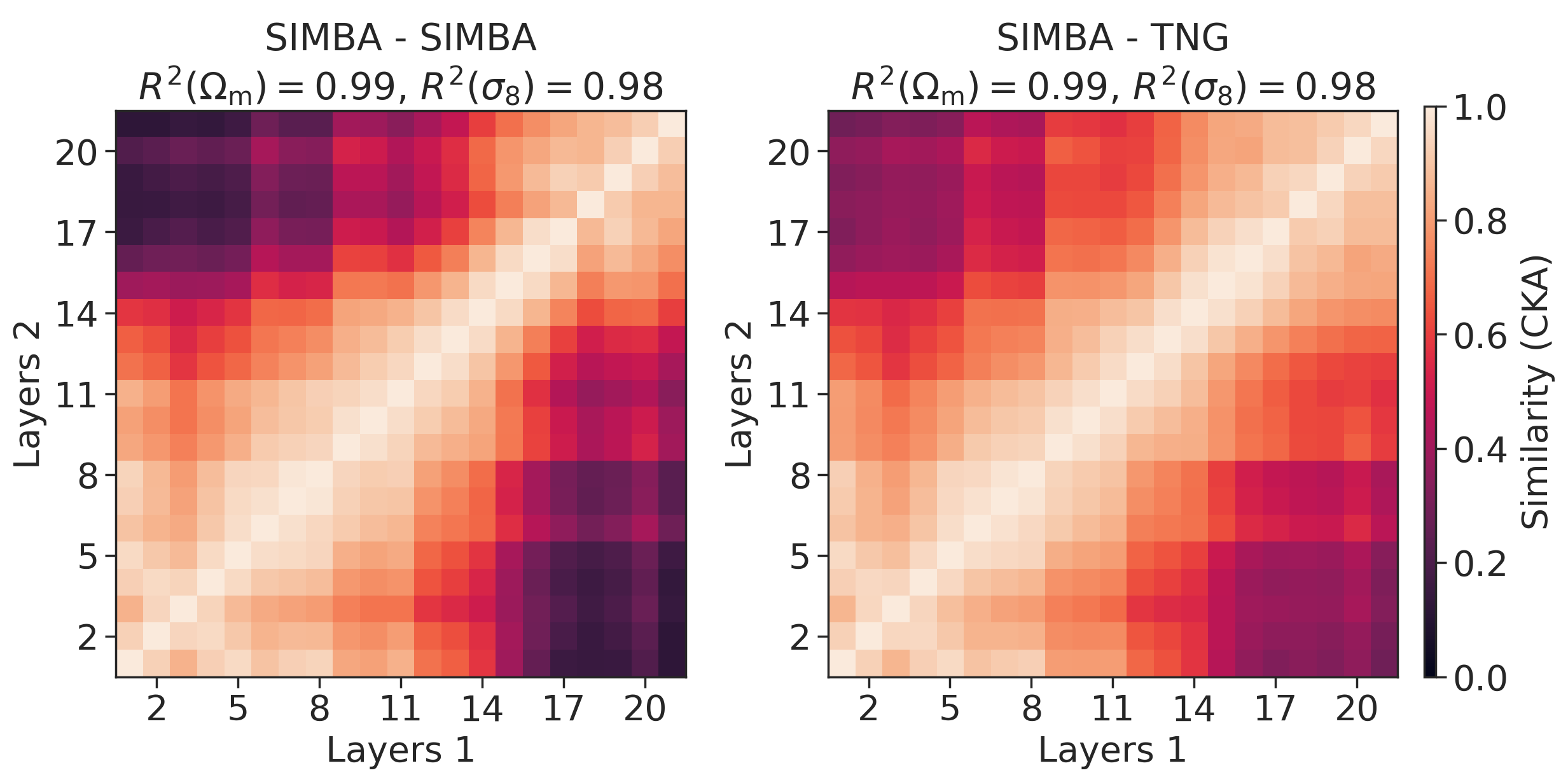}
    \caption{CKA matrices for the ID ({\sc SIMBA-SIMBA}) (left) and OOD ({\sc SIMBA-TNG}) (right) cases. The titles show $R^2$ scores between model prediction and true value for the $\Omega_{\rm m}$ and $\sigma_8$ cosmological parameters.}\label{fig:mtot_cka}
  \end{subfigure}
  
  \vspace{3mm}
  \begin{subfigure}[t]{\textwidth}
    \centering
    \includegraphics[width=\textwidth,keepaspectratio]{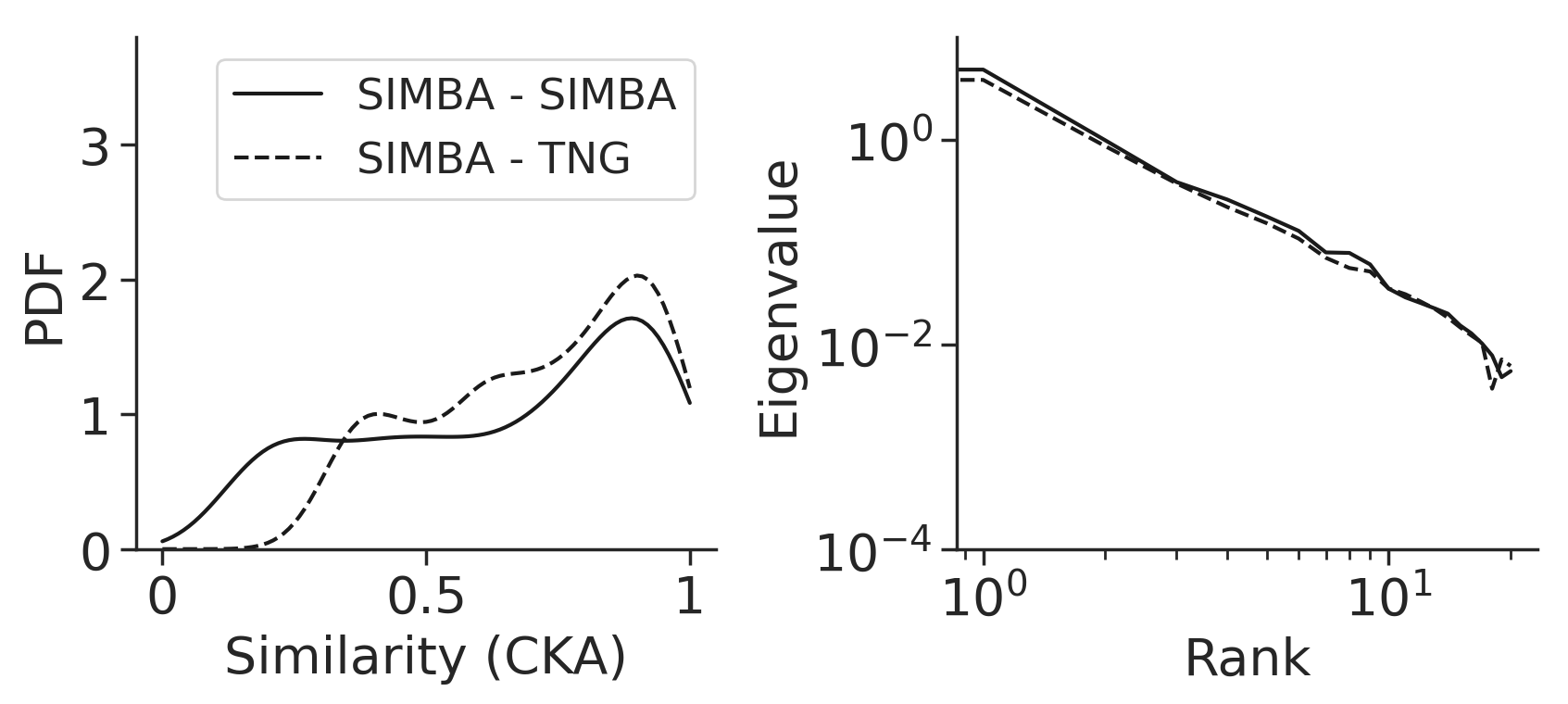}
    \caption{Summary statistics of the CKA matrices from Fig.~\ref{fig:mtot_cka}. On the left is the probability density function (PDF) of the CKA similarities, and on the right is the eigenvalues as a function of rank (i.e., the spectrum).}\label{fig:mtot_stats}
  \end{subfigure}
  \caption{Analysis of the CKA similarities for the Mtot field. Representations of layers of the CNN trained on the {\sc SIMBA} simulations are diverse not only when the CNN is tested on 2D maps from {\sc SIMBA} (i.e., ID samples) but also when tested on 2D maps from TNG (i.e., OOD samples). The $R^2$ scores for {\sc SIMBA-TNG} are the same as {\sc SIMBA-SIMBA}, suggesting that the model robustly generalizes to OOD ({\sc TNG}) data.}\label{fig:Mtot}
  \end{minipage}
\end{figure*}

\subsection{When a model fails to generalize OOD, outputs of different layers are similar}

For the Temperature field, the model fails to generalize from the training simulation environment, TNG, to the OOD simulation environment, SIMBA. This failure to generalize is because while $R^2$ scores for the cosmological parameters ($\Omega_m$, $\sigma_8$) for {\sc TNG--TNG} are high (0.99 for $\Omega_m$, 0.98 for $\sigma_8$), they deteriorate for {\sc TNG--SIMBA} (0.16 for $\Omega_m$, -10.61 for $\sigma_8$).

The CKA matrices for {\sc TNG--TNG} and {\sc TNG--SIMBA} shown in Fig.~\ref{fig:temp_cka} show a perfect similarity (= 1.0) along the main diagonal (i.e., bottom-left to top-right of the CKA matrix), which is expected since the scores on the diagonal compare each layer's representation with itself. However, the similarities in the non-diagonal entries for the ID setting {\sc TNG--TNG} are much smaller than those of the OOD setting {\sc TNG--SIMBA}, which exhibits a block structure. The block structures for {\sc TNG--SIMBA} indicate that different layers of the model produce similar representations on OOD inputs. In other words, the representations change only slightly as the OOD samples evolve deeper into this network.

We also quantify the differences in off-diagonal similarities in the CKA matrix that use ID and OOD data. The summary statistics (Fig.~\ref{fig:temp_stats}) empirically show that the following hold for CKA matrices containing block structures compared to matrices without dominant block or diffused structures: (a) the distribution of the CKA similarities is more skewed towards larger values, and (b) their eigenvalues are smaller. The latter implies that the high CKA similarities are dispersed throughout the matrix rather than being dominantly focused on the diagonal, leading to smaller eigenvalues.

\subsection{When a model successfully generalizes OOD, outputs of different layers are dissimilar}
In the previous section, we illustrated how CKA similarities change between ID and OOD data for models that generalize poorly between simulations. We repeat the CKA analysis on models trained on the Mtot maps, which can successfully generalize between simulations. For the Mtot field, the $R^2$ scores for the cosmological parameters ($\Omega_m$, $\sigma_8$) in the ID setting {\sc SIMBA--SIMBA} are high (0.99 for $\Omega_m$, 0.98 for  $\sigma_8$). The $R^2$ scores remain identical for the OOD test setting {\sc SIMBA--TNG} (0.99 for $\Omega_m$, 0.98 for  $\sigma_8$). Therefore, the model is robust to the change in the simulation environment, perhaps partly because the shift appears smaller (Fig.~\ref{fig:mtot_ex}). 

In contrast to Temperature, the CKA matrices for ID ({\sc SIMBA--SIMBA}) and OOD ({\sc SIMBA--TNG}) are similar visually (Fig.~\ref{fig:mtot_cka}). Although the block structures in the CKA matrix may not have been entirely removed for {\sc SIMBA--TNG}, the representations of initial (1--5) and final (16--20) layers have become less similar than in the CKA of the OOD setting of Temperature ({\sc TNG--SIMBA}; see Fig.~\ref{fig:temp_cka}). In the Mtot setting, therefore, representations at the final layers have been substantially modified compared to those in the initial layers, standing in stark contrast to the behavior of a model during an OOD generalization failure as seen in Fig.~\ref{fig:temp_cka}. 

Unlike Fig.~\ref{fig:temp_stats}, the summary statistics in Fig.~\ref{fig:mtot_stats} show that the eigenvalues and the distribution of the CKA similarities are similar for ID and OOD cases. These statistics quantify the visual differences between CKA matrices containing block structures and those that do not contain block structures. Overall, our experiments show that models produce different outputs at each layer when successfully generalizing OOD, but not when failing to generalize.

\section{Conclusion and future work}
Studying the CKA matrices of pre-trained CNNs tested using ID and OOD samples, we find that poor test-time model accuracy corresponds to higher similarity between different layers of the model (non-diagonal entries in the CKA matrix).

A high-similarity block feature in the CKA matrix suggests that these layers are unnecessary, and a similar accuracy can be obtained by replacing these layers with a single layer and then performing the training. In future work, therefore, the CKA matrix can be used to prune layers that correspond to diffused or block structures, which could reduce the memory footprint of the models while maintaining similar test-time performance. In addition, we have found that robust models have a characteristic feature of modifying their representations across different layers, whereas non-robust models possess stagnancy in their representations across different layers. This observation is crucial to the OOD generalization problem.  

Our approach may be used to decide what simulations are best to train models for application on real-world data. By identifying when the model is not robust to a distribution shift, one can select data most similar to the OOD setting and produce novel training datasets capturing distribution shifts. Identifying generalization failures in this way is one method to build generally robust models, serving as feedback for improving training strategies to achieve OOD generalization. By understanding model behavior when failing to generalize, we can measure model confidence at test time and identify problem areas for learned models. 


Future work can also investigate the causal impact of similarity structures on generalization performance in more detail. We plan to optimize these models by including the CKA matrix, as inductive bias, in the loss function to enforce similarity between CKA matrices of the ID and OOD samples, hence achieving OOD generalization.
The advantage of this approach is that the computation of the CKA matrix does not require knowledge of physical parameters. This mimics the scenario of extracting information from real observations, where physical parameters are always unknown. 



\begin{ack}
The authors acknowledge helpful discussions with Benjamin D. Wandelt and Francisco Villaescusa-Navarro. SH acknowledges support for Program number HST-HF2-51507 provided by NASA through a grant from the Space Telescope Science Institute, which is operated by the Association of Universities for Research in Astronomy, incorporated, under NASA contract NAS5-26555. SH also acknowledges support from the NYU Office of Postdoctoral Affairs and Simons Foundation. This work was supported by Hyundai Motor Company (under the project Uncertainty in Neural Sequence Modeling) and the Samsung Advanced Institute of Technology (under the project Next Generation Deep Learning: From Pattern Recognition to AI). This work has been made possible in part by a gift from the Chan Zuckerberg Initiative Foundation to establish the Kempner Institute for the Study of Natural and Artificial Intelligence. SA acknowledges financial support from the South African Radio Astronomy Observatory (SARAO). 
\end{ack}


\bibliographystyle{unsrtnat}
\bibliography{biblio}

\begin{thebibliography}{12}
\providecommand{\natexlab}[1]{#1}
\providecommand{\url}[1]{\texttt{#1}}
\expandafter\ifx\csname urlstyle\endcsname\relax
  \providecommand{\doi}[1]{doi: #1}\else
  \providecommand{\doi}{doi: \begingroup \urlstyle{rm}\Url}\fi

\bibitem[{Ntampaka} et~al.(2021){Ntampaka}, {Ho}, and
  {Nord}]{2021arXiv211114566N}
Michelle {Ntampaka}, Matthew {Ho}, and Brian {Nord}.
\newblock {Building Trustworthy Machine Learning Models for Astronomy}.
\newblock \emph{arXiv e-prints}, art. arXiv:2111.14566, November 2021.
\newblock \doi{10.48550/arXiv.2111.14566}.

\bibitem[Matilla et~al.(2020)Matilla, Sharma, Hsu, and
  Haiman]{PhysRevD.102.123506}
Jos\'e Manuel~Zorrilla Matilla, Manasi Sharma, Daniel Hsu, and Zolt\'an Haiman.
\newblock Interpreting deep learning models for weak lensing.
\newblock \emph{Phys. Rev. D}, 102:\penalty0 123506, Dec 2020.
\newblock \doi{10.1103/PhysRevD.102.123506}.
\newblock URL \url{https://link.aps.org/doi/10.1103/PhysRevD.102.123506}.

\bibitem[Morice-Atkinson et~al.(2018)Morice-Atkinson, Hoyle, and
  Bacon]{10.1093/mnras/sty2575}
Xan Morice-Atkinson, Ben Hoyle, and David Bacon.
\newblock {Learning from the machine: interpreting machine learning algorithms
  for point- and extended-source classification}.
\newblock \emph{Monthly Notices of the Royal Astronomical Society},
  481\penalty0 (3):\penalty0 4194--4205, 09 2018.
\newblock ISSN 0035-8711.
\newblock \doi{10.1093/mnras/sty2575}.
\newblock URL \url{https://doi.org/10.1093/mnras/sty2575}.

\bibitem[Wu(2020)]{Wu_2020}
John~F. Wu.
\newblock Connecting optical morphology, environment, and h i mass fraction for
  low-redshift galaxies using deep learning.
\newblock \emph{The Astrophysical Journal}, 900\penalty0 (2):\penalty0 142, sep
  2020.
\newblock \doi{10.3847/1538-4357/abacbb}.
\newblock URL \url{https://dx.doi.org/10.3847/1538-4357/abacbb}.

\bibitem[{Cranmer}(2023)]{2023arXiv230501582C}
Miles {Cranmer}.
\newblock {Interpretable Machine Learning for Science with PySR and
  SymbolicRegression.jl}.
\newblock \emph{arXiv e-prints}, art. arXiv:2305.01582, May 2023.
\newblock \doi{10.48550/arXiv.2305.01582}.

\bibitem[{Cianfarani} et~al.(2022){Cianfarani}, {Nitin Bhagoji}, {Sehwag},
  {Zhao}, {Mittal}, and {Zheng}]{2022arXiv220609868C}
Christian {Cianfarani}, Arjun {Nitin Bhagoji}, Vikash {Sehwag}, Ben~Y. {Zhao},
  Prateek {Mittal}, and Haitao {Zheng}.
\newblock {Understanding Robust Learning through the Lens of Representation
  Similarities}.
\newblock \emph{arXiv e-prints}, art. arXiv:2206.09868, June 2022.
\newblock \doi{10.48550/arXiv.2206.09868}.

\bibitem[{Villaescusa-Navarro} et~al.(2021{\natexlab{a}}){Villaescusa-Navarro},
  {Genel}, {Angles-Alcazar}, {Thiele}, {Dave}, {Narayanan}, {Nicola}, {Li},
  {Villanueva-Domingo}, {Wandelt}, {Spergel}, {Somerville}, {Zorrilla Matilla},
  {Mohammad}, {Hassan}, {Shao}, {Wadekar}, {Eickenberg}, {Wong}, {Contardo},
  {Jo}, {Moser}, {Lau}, {Machado Poletti Valle}, {Perez}, {Nagai}, {Battaglia},
  and {Vogelsberger}]{CMD2021}
Francisco {Villaescusa-Navarro}, Shy {Genel}, Daniel {Angles-Alcazar}, Leander
  {Thiele}, Romeel {Dave}, Desika {Narayanan}, Andrina {Nicola}, Yin {Li},
  Pablo {Villanueva-Domingo}, Benjamin {Wandelt}, David~N. {Spergel}, Rachel~S.
  {Somerville}, Jose~Manuel {Zorrilla Matilla}, Faizan~G. {Mohammad}, Sultan
  {Hassan}, Helen {Shao}, Digvijay {Wadekar}, Michael {Eickenberg}, Kaze W.~K.
  {Wong}, Gabriella {Contardo}, Yongseok {Jo}, Emily {Moser}, Erwin~T. {Lau},
  Luis~Fernando {Machado Poletti Valle}, Lucia~A. {Perez}, Daisuke {Nagai},
  Nicholas {Battaglia}, and Mark {Vogelsberger}.
\newblock {The CAMELS Multifield Dataset: Learning the Universe's Fundamental
  Parameters with Artificial Intelligence}.
\newblock \emph{arXiv e-prints}, art. arXiv:2109.10915, September
  2021{\natexlab{a}}.

\bibitem[{Villaescusa-Navarro} et~al.(2021{\natexlab{b}}){Villaescusa-Navarro},
  {Angl{\'e}s-Alc{\'a}zar}, {Genel}, {Spergel}, {Somerville}, {Dave},
  {Pillepich}, {Hernquist}, {Nelson}, {Torrey}, {Narayanan}, {Li}, {Philcox},
  {La Torre}, {Maria Delgado}, {Ho}, {Hassan}, {Burkhart}, {Wadekar},
  {Battaglia}, {Contardo}, and {Bryan}]{CAMELS}
Francisco {Villaescusa-Navarro}, Daniel {Angl{\'e}s-Alc{\'a}zar}, Shy {Genel},
  David~N. {Spergel}, Rachel~S. {Somerville}, Romeel {Dave}, Annalisa
  {Pillepich}, Lars {Hernquist}, Dylan {Nelson}, Paul {Torrey}, Desika
  {Narayanan}, Yin {Li}, Oliver {Philcox}, Valentina {La Torre}, Ana {Maria
  Delgado}, Shirley {Ho}, Sultan {Hassan}, Blakesley {Burkhart}, Digvijay
  {Wadekar}, Nicholas {Battaglia}, Gabriella {Contardo}, and Greg~L. {Bryan}.
\newblock {The CAMELS Project: Cosmology and Astrophysics with Machine-learning
  Simulations}.
\newblock \emph{\apj}, 915\penalty0 (1):\penalty0 71, July 2021{\natexlab{b}}.
\newblock \doi{10.3847/1538-4357/abf7ba}.

\bibitem[{Kornblith} et~al.(2019){Kornblith}, {Norouzi}, {Lee}, and
  {Hinton}]{2019arXiv190500414K}
Simon {Kornblith}, Mohammad {Norouzi}, Honglak {Lee}, and Geoffrey {Hinton}.
\newblock {Similarity of Neural Network Representations Revisited}.
\newblock \emph{arXiv e-prints}, art. arXiv:1905.00414, May 2019.
\newblock \doi{10.48550/arXiv.1905.00414}.

\bibitem[Nguyen et~al.(2021)Nguyen, Raghu, and Kornblith]{nguyen2021wide}
Thao Nguyen, Maithra Raghu, and Simon Kornblith.
\newblock Do wide and deep networks learn the same things? uncovering how
  neural network representations vary with width and depth, 2021.

\bibitem[{Villaescusa-Navarro} et~al.(2021{\natexlab{c}}){Villaescusa-Navarro},
  {Angl{\'e}s-Alc{\'a}zar}, {Genel}, {Spergel}, {Li}, {Wandelt}, {Nicola},
  {Thiele}, {Hassan}, {Zorrilla Matilla}, {Narayanan}, {Dave}, and
  {Vogelsberger}]{2021arXiv210909747V}
Francisco {Villaescusa-Navarro}, Daniel {Angl{\'e}s-Alc{\'a}zar}, Shy {Genel},
  David~N. {Spergel}, Yin {Li}, Benjamin {Wandelt}, Andrina {Nicola}, Leander
  {Thiele}, Sultan {Hassan}, Jose~Manuel {Zorrilla Matilla}, Desika
  {Narayanan}, Romeel {Dave}, and Mark {Vogelsberger}.
\newblock {Multifield Cosmology with Artificial Intelligence}.
\newblock \emph{arXiv e-prints}, art. arXiv:2109.09747, September
  2021{\natexlab{c}}.
\newblock \doi{10.48550/arXiv.2109.09747}.

\bibitem[Villaescusa-Navarro et~al.(2022)Villaescusa-Navarro, Genel,
  Anglés-Alcázar, Thiele, Dave, Narayanan, Nicola, Li, Villanueva-Domingo,
  Wandelt, Spergel, Somerville, Matilla, Mohammad, Hassan, Shao, Wadekar,
  Eickenberg, Wong, Contardo, Jo, Moser, Lau, Valle, Perez, Nagai, Battaglia,
  and Vogelsberger]{Villaescusa-Navarro_2022}
Francisco Villaescusa-Navarro, Shy Genel, Daniel Anglés-Alcázar, Leander
  Thiele, Romeel Dave, Desika Narayanan, Andrina Nicola, Yin Li, Pablo
  Villanueva-Domingo, Benjamin Wandelt, David~N. Spergel, Rachel~S. Somerville,
  Jose Manuel~Zorrilla Matilla, Faizan~G. Mohammad, Sultan Hassan, Helen Shao,
  Digvijay Wadekar, Michael Eickenberg, Kaze W.~K. Wong, Gabriella Contardo,
  Yongseok Jo, Emily Moser, Erwin~T. Lau, Luis Fernando Machado~Poletti Valle,
  Lucia~A. Perez, Daisuke Nagai, Nicholas Battaglia, and Mark Vogelsberger.
\newblock The camels multifield data set: Learning the universe’s fundamental
  parameters with artificial intelligence.
\newblock \emph{The Astrophysical Journal Supplement Series}, 259\penalty0
  (2):\penalty0 61, apr 2022.
\newblock \doi{10.3847/1538-4365/ac5ab0}.
\newblock URL \url{https://dx.doi.org/10.3847/1538-4365/ac5ab0}.

\end{thebibliography}







\end{document}